 \documentclass[final,authoryear,5p,times,twocolumn]{elsarticle}

 \usepackage{graphicx}
\usepackage{amssymb}
\journal{IJA}

\begin{document}

\begin{frontmatter}

%% Title, authors and addresses

%% use the tnoteref command within \title for footnotes;
%% use the tnotetext command for the associated footnote;
%% use the fnref command within \author or \address for footnotes;
%% use the fntext command for the associated footnote;
%% use the corref command within \author for corresponding author footnotes;
%% use the cortext command for the associated footnote;
%% use the ead command for the email address,
%% and the form \ead[url] for the home page:
%%
%% \title{Title\tnoteref{label1}}
%% \tnotetext[label1]{}
%% \author{Name\corref{cor1}\fnref{label2}}
%% \ead{email address}
%% \ead[url]{home page}
%% \fntext[label2]{}
%% \cortext[cor1]{}
%% \address{Address\fnref{label3}}
%% \fntext[label3]{}

\title{Towards a comprehensive model of Earth's disk-integrated Stokes vector}

%% use optional labels to link authors explicitly to addresses:
%% \author[label1,label2]{<author name>}
%% \address[label1]{<address>}
%% \address[label2]{<address>}

\author{A. Garc\'ia Mu\~noz}

\address{ESA Fellow, ESA/RSSD, ESTEC, 2201 AZ Noordwijk, the Netherlands}
\ead{tonhingm@gmail.com}

%\author{F.P. Mills}

%\address{Research School of Physics and Engineering and Fenner School of Environment and Society, Australian National University,
%Canberra, ACT 0200, Australia}
%\address{Space Science Institute, Boulder, CO 80301, USA}
%\ead{frank.mills@anu.edu.au}

\begin{abstract}
%% Text of abstract

A significant body of work on simulating the remote appearance of Earth-like 
exoplanets has been done over the last decade. 
The research is driven by the prospect of characterizing habitable
planets beyond the Solar System in the near future. 
In this work, I present a method to produce 
the disk-integrated signature of planets that are described in their three-dimensional
complexity, \textit{i.e.} with both horizontal and vertical variations in the
optical properties of their envelopes. 
The approach is based on Pre-conditioned Backward Monte Carlo integration of the
vector Radiative Transport Equation and 
yields the full Stokes vector for outgoing reflected radiation. 
The method is demonstrated through selected examples
inspired by published work at wavelengths from 
the visible to the near infrared and terrestrial prescriptions 
of both cloud and surface albedo maps. 
I explore the performance of the method in terms of computational time and accuracy. 
A clear advantage of this approach is that its computational cost does not appear
to be significantly affected by non-uniformities in the planet optical properties.
Earth's simulated appearance is strongly dependent on wavelength; 
both brightness and polarisation undergo diurnal variations arising from changes in the planet cover,
but polarisation yields a better insight into variations with phase angle. 
There is partial cancellation of the polarised signal from the 
northern and southern hemispheres so that the outgoing polarisation vector lies preferentially 
either in the plane parallel or perpendicular to the planet scattering plane, 
also for non-uniform cloud and albedo properties and various levels of absorption 
within the atmosphere.
The evaluation of circular polarisation is challenging; a
number of one-photon experiments of 10$^9$ or more is needed to resolve
hemispherically-integrated degrees of circular polarisation of a few times 10$^{-5}$. 
Last, I introduce brightness curves of Earth 
obtained with one of the Messenger cameras at three wavelengths (0.48, 0.56 
and 0.63 $\mu$m)
during a flyby in 2005. 
The light curves show distinct structure associated with
the varying aspect of the Earth's visible disk (phases of 98--107$^{\circ}$) as the planet undergoes a 
full 24h rotation; the structure is reasonably well reproduced with model simulations.

\end{abstract}

\begin{keyword}
%% keywords here, in the form: keyword \sep keyword

%% MSC codes here, in the form: \MSC code \sep code
%% or \MSC[2008] code \sep code (2000 is the default)

\end{keyword}

\end{frontmatter}

%\linenumbers

%% main text

\clearpage
\section{\label{introduction_sec}Introduction}

Our understanding of exoplanetary atmospheres relies on the remote sensing of radiation that,
arising from the host star or the planet itself, becomes imprinted with some of the 
planet's atmospheric features. Ongoing technological developments, to be implemented in 
$>$30-m ground-based telescopes and dedicated space missions, 
are steadily increasing the number
of exoplanets amenable to atmospheric characterisation. In the foreseeable future, 
sophisticated instruments will allow us to separate the radiation of an Earth
twin from the glare of its host star \citep[e.g.][and refs. therein]{trauboppenheimer2010}. At that moment, 
it will be possible to address questions such as the occurrence of life on the planet by
searching for selected biosignatures in its reflected and/or emitted spectrum
\citep[e.g.][]{brandtspiegel2014,desmaraisetal2002,seageretal2005,sparksetal2009}

In preparation for that moment, modellers have been setting up and testing the tools with
which one day the (one pixel, at first) images of Earth-like exoplanets will be rationalized 
\citep[e.g.][]{fordetal2001,karalidietal2011,karalidietal2012,karalidistam2012,robinsonetal2011,stam2008,tinettietal2006,zuggeretal2010,zuggeretal2011}. 
As some of those works have shown, the information contained in spectra
and colour photometry of the one-pixel images will inform us about aspects of the planet 
such as its atmospheric composition, existence of clouds, 
and land/ocean partitioning. A key aspect of models, 
and one that is investigated here,
is their capacity to predict an exoplanet's remote appearance and, in particular,
how to treat the three-dimensional nature of the atmospheres. 
The need for such a treatment becomes more apparent as a number of groups 
are beginning to explore
the coupled effects of atmospheric dynamics, heat redistribution and chemistry
in the framework of General Circulation Models 
\citep[e.g.][]{,caroneetal2014,joshi2003,katariaetal2014,menou2012,zaluchaetal2013}.

Modelling radiative transport in three-dimensional atmospheres is a computationally 
intensive task. For that reason it is important to explore all possible avenues. 
The two most usual approaches to the problem are:
%\begin{itemize}
\begin{enumerate}
\item The planet disk is partitioned into an $n_x$$\times$$n_y$ number of disk elements. 
The outgoing radiation is calculated at each discrete element 
under the approximation of locally 
plane-parallel atmosphere; then, each discrete contribution is properly added up to generate the disk-integrated magnitude. 
%This has been the preferred approach so far
%\citep{karalidistam2012,karalidietal2012,tinettietal2006}. 
Its computational cost is proportional to $n_x$$\times$$n_y$. 
As an example, simulating 
a fully-illuminated planet at the 2$^{\circ}$$\times$2$^{\circ}$
resolution level involves 8100 separate calculations. 
\item Disk integration is directly approached  by means of Monte Carlo (MC)
integration. The MC scheme conducts one-photon numerical experiments, each of them 
tracing a photon trajectory through the atmospheric medium. 
Repeating the experiment $n_{\rm{ph}}$
times results in an estimate of the disk-integrated outgoing radiation 
that converges to the exact magnitude with a standard deviation that follows a
$\sim$$n^{-1/2}_{\rm{ph}}$ law. \citet{fordetal2001} utilised a Forward
MC algorithm to investigate the diurnal variability of the Earth's brightness. 
Forward refers to the tracing of the photons starting from the radiation source
onwards through the atmospheric medium, \textit{i.e.} in the same order as the 
actual photon displacements. 
In the MC approach, the total computational cost is proportional to $n_{\rm{ph}}$. 
\end{enumerate}
%\end{itemize}

Recent work has introduced an alternative approach based on Pre-conditioned 
Backward Monte Carlo (PBMC) integration of the vector Radiative Transport Equation
\citep{garciamunozmills2014}. 
Past applications of the PBMC model to the problem of scattering from 
a spatially-unresolved planet focused on atmospheres that might be stratified in the 
vertical but exhibited otherwise uniform optical properties in the horizontal direction.
The importance of accounting for three-dimensional effects in the description of the
optical properties of exoplanets has been substantiated by a number of works 
\citep{fordetal2001,karalidistam2012,karalidietal2012,tinettietal2006}.

Here, I investigate further the PBMC approach and use it to explore 
Earth's remote appearance for non-uniform distributions of cloud and surface properties. 
The methodology may find application in 
the interpretation of earthshine observations 
\citep[e.g.][]{arnoldetal2002,bazzonetal2013,sterziketal2012,woolfetal2002} and 
in related research fields
such as climate studies that are concerned
with planet-averaged properties. 
This is ongoing work that demonstrates the potentiality of the
PBMC approach to become a reference technique in the production of disk-integrated curves
for exoplanets.

Section \S\ref{PBMC_sec} introduces the method, 
\S\ref{atmosphere_section} describes the 
atmospheric, cloud and surface properties implemented in the current 
exploration exercise, and 
\S\ref{results_section} comments on the phase and diurnal light
curves of Earth for a few wavelengths and two different cloud covers. 
Finally, section \S\ref{messenger_section} introduces the 
brightness light curves of Earth
obtained with images from the 2005 Earth flyby of the Messenger spacecraft,  
and \S\ref{outlook_section} summarises the main conclusions and 
touches upon follow-up work.

\clearpage
\section{\label{PBMC_sec} The PBMC algorithm}

The fundamentals of the PBMC algorithm have been described \citep{garciamunozmills2014}, and
the model has been validated for horizontally-uniform atmospheres
\citep{garciamunozmills2014, garciamunozetal2014}.
The validation exercise included a few thousand solutions from
the literature for plane-parallel media, disk-integrated
model solutions for Rayleigh-scattering atmospheres and 
measurements of the disk-integrated brightness 
and polarisation of Venus. 
In the comparison against other model calculations, typical accuracies of 0.1\% and better
are achieved.

The algorithm calculates the stellar radiation reflected from the whole disk for
planets with stratified atmospheres. 
The calculation 
proceeds by simulating a sequence of $n_{\rm{ph}}$ one-photon numerical 
experiments. 
Being a backward algorithm, 
each experiment traces a photon trajectory from the observer through 
the scattering medium. The contribution to the estimated radiation as
measured by the observer is 
built up by partial contributions every time the photon undergoes a collision. 
The simulation stops when the photon is either fully absorbed in or escapes from the
atmosphere. 
Pre-conditioning refers to a novel scheme 
that incorporates the polarisation history of the 
photon (since its departure from the observer's location) 
into the sampling of propagation directions. 
Pre-conditioning prevents spurious behaviours 
in conservative, optically-thick, strongly-polarising media, and 
accelerates the convergence over the non-pre-conditioned treatment.

The algorithm includes a 
scheme to select the photon entry point into 
the atmosphere based on the local projected area of the 'visible' planet disk
\citep{garciamunozmills2014}.  
This strategy ensures that all simulated photons contribute to the estimated 
radiation at the observer's location. 
In this approach, each solution to the radiative problem 
is specific to the details of the illuminated disk as viewed by
the observer. Changing that view requires a new radiative transport 
calculation.
The implementation is based on the vector Radiative Transport Equation, 
which entails that the model computes the disk-integrated Stokes vector of the
planet. 
Not explored in past applications of the PBMC approach, 
the model is well suited for investigating planets with
non-uniform distributions of clouds and albedos.

%% The Appendices part is started with the command \appendix;
%% appendix sections are then done as normal sections
%% \appendix

\clearpage
\section{\label{atmosphere_section}Atmospheric and surface model}

The PBMC algorithm handles elaborate 
descriptions of the atmospheric, cloud and surface optical properties of the planet. 
To keep the number of parameters reasonable, however, it is convenient to introduce various
simplifications, which are described in what follows.

The model atmosphere adopted here
includes gas and clouds and utilises a 30-slab vertical grid with
resolutions of 1, 2 and 4 km for altitudes in the 
0--10, 10--30 and 30--70 km ranges. 
The gas is stratified with a scale height of 8 km and the total
Rayleigh optical thickness is parameterised
by $\tau_{\rm{g}}$=0.1(0.555/$\lambda$ [$\mu$m])$^4$. 
Clouds are placed in a single slab and assumed to be composed of liquid 
water and have 
wavelength-dependent real refractive indices ranging  
from 1.34 at 0.47 $\mu$m to 1.30 at 2.13 $\mu$m. 
The considered cloud droplet sizes have an
effective radius and variance of 
$r_{\rm{eff}}$=7.33 $\mu$m and $v_{\rm{eff}}$=0.12, respectively, which is
appropriate for continental stratus clouds \citep{hessetal1998}.
The scattering matrix for the polarisation calculations is determined 
from Mie theory \citep{mishchenkoetal2002}. 
The cloud optical thickness $\tau_{\rm{c}}$ is assumed to be the same for 
all clouds on the planet. 
The model atmosphere includes absorption by ozone 
in the Chappuis band \citep{jacquinethussonetal2005}.
The ozone vertical profile 
peaks in the stratosphere at 25 km and has an integrated column of 275
Dobson units \citep{loughmanetal2004}.
Ozone absorption is stronger between 0.55 and 0.65 $\mu$m, and its effect on the
radiative transport calculations is particularly distinct in that spectral range.
No other gas molecular bands were considered, and therefore the study focuses
on the spectral continuum  rather than on strong absorption molecular features. 
Future work will investigate the potential advantages of introducing the 
spectral direction into the Monte Carlo integration. Foreseeably, 
associating each one-photon experiment with a properly-selected wavelength (over an
instrument's bandpass) and with a photon entry point into the atmosphere
may lead to further computational savings in the simulation of spectra with respect
to the usual strategy of stacking many monochromatic calculations 
and subsequently degrading the so-formed spectrum. 
The savings are likely to be significant if the final spectra 
are only required at moderate-to-low resolving powers.

At the planet surface, the model adopts the MODIS white-sky albedos 
specific to early
August 2004 \citep{moodyetal2005}.\footnote{Accessible through 
http://modis-atmos.gsfc.nasa.gov/ALBEDO/} 
The albedos are available at bands centered at
0.47, 0.555, 0.659, 0.858, 1.24, 1.64 and 2.13 $\mu$m,
which cover much of the spectrum dominated by reflected radiation.
The bandwidths of the MODIS filters are of 20--50 nm \citep{xiongetal2009}. 
The monochromatic model calculations presented here are carried out at the
MODIS central wavelengths.
The prescribed albedos are identically zero over the ocean. 
Surface reflection is throughout assumed to be of Lambertian type. 
Specular reflection at the ocean,  
which may contribute to the planet signal at large
phase angles \citep{robinsonetal2010, williamsgaidos2008,zuggeretal2010,zuggeretal2011}, 
will be implemented in later work.

The adopted maps of cloud fraction are from MODIS data\footnote{ 
Accessible through the MODIS Terra and Aqua Daily Level-3 Data site,\\
http://gdata1.sci.gsfc.nasa.gov/daac-bin/G3/gui.cgi?instance$\_$id=MODIS$\_$DAILY$\_$L3} specific to 3rd August 2005  
(see Section \S\ref{messenger_section}). 
For each one-photon experiment, a scheme based on the local cloud fraction
$f_{\rm{c}}$ at the photon entry point into the atmosphere
determines the surface albedo to be utilised in that specific photon trajectory
simulation and whether the photon is traced through a gas-only or gas-plus-cloud
medium.
Internally, the scheme draws a random number $\rho_{\rm{c}}$$\in$[0, 1]. 
If $\rho_{\rm{c}}$$\le$$f_{\rm{c}}$, the photon is traced through 
an atmosphere of optical thickness $\tau_{\rm{g}}$+$\tau_{\rm{c}}$ 
with optical properties within each slab 
properly averaged over the gas and cloud. Otherwise, if  
$\rho_{\rm{c}}$$>$$f_{\rm{c}}$, the photon is traced through a gas-only atmosphere of
optical thickness $\tau_{\rm{g}}$. 

\begin{figure*}
\begin{centering}
\noindent\includegraphics[width=9cm]{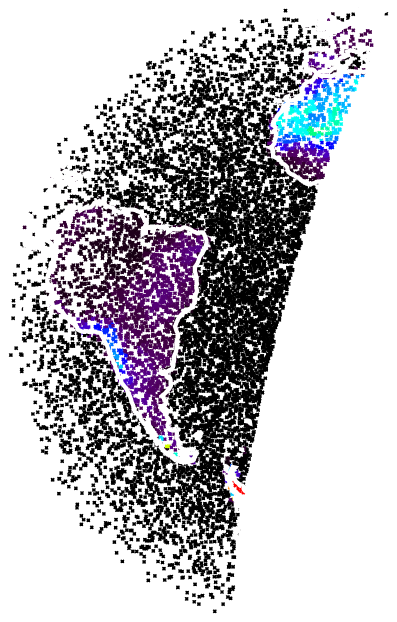}
\\
\vspace{-3.00cm} 
\noindent\includegraphics[width=9cm]{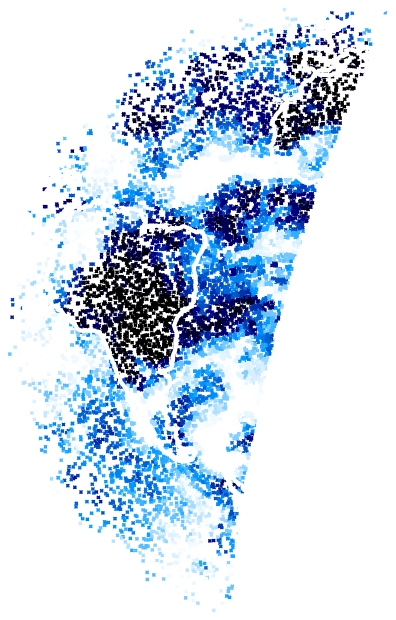}
\caption{\label{albedocloudmap_fig} Top. In the PBMC algorithm, each one-photon experiment is
associated with a local albedo value. This image shows the surface albedo 
at 0.47 $\mu$m as sampled in a
$n_{\rm{ph}}$=10$^4$ calculation for viewing and illumination conditions specific to
the Messenger view of Earth on 18:30UT 3rd August 2005. 
The phase angle for this view of the Earth is equal to 98.7$^{\circ}$.
The density of sampling 
points is based on the projected area of the planet's visible disk, 
which explains why the sampling points tend to concentrate near the terminator.
Bottom. Same as above 
but showing the cloud fraction $f_c$. 
White and black stand for $f_c$=1 and =0 respectively.
} 
\end{centering}
\end{figure*}

Both the  input surface albedo and cloud fraction maps 
are averaged and mapped onto longitude/latitude maps of
2$^{\circ}$$\times$2$^{\circ}$ resolution. 
Each map is read into the model at the beginning of the run 
and kept in memory 
throughout the simulation of the $n_{\rm{ph}}$ one-photon experiments.
Figure (\ref{albedocloudmap_fig}) shows albedo and cloud fraction maps 
as sampled by the model in a run
specific to the Messenger view of Earth on 18:30UT 3rd August 2005.

\newpage
\section{\label{results_section}Model results}
\subsection{\label{convergence_section}A look into convergence}

In the implementation of the PBMC algorithm, the irradiance $\mathbf{F}$
at the observer's location is evaluated through a summation:
\begin{equation}
\mathbf{F}= \frac{\pi}{2}(1+\cos{(\alpha)}) 
\frac{1}{n_{\rm{ph}}} \sum_{i=1}^{n_{\rm{ph}}} \mathbf{I}(u_i,v_i),
\label{summation_eq}
\end{equation} 
where $\mathbf{I}(u_i,v_i)$ is the outgoing radiance Stokes vector 
from the $i$-th one-photon experiment, the uniform random variables 
$u$, $v$$\in$[0, 1] sample the planet visible disk
and $\alpha$ is the star-planet-observer phase angle
\citep{garciamunozmills2014}. 
Equation (\ref{summation_eq}) is effectively an arithmetic average and 
its convergence properties 
depend on the dispersion in the outcome of the one-photon experiments. 
$F_I$ is the first element of $\mathbf{F}$ and in the adopted normalisation 
it is identical to 
$A_g$$\Phi$($\alpha$), where $A_g$ is the planet's geometric albedo and 
$\Phi$($\alpha$) ($\Phi$(0)$\equiv$1) is the planet's scattering phase function. 
Properly, the planet's scattering phase function 
also depends on the optical properties of the planet as viewed from the observer's 
location, and they may change over time.

To investigate the convergence properties of $\mathbf{F}$, 
it is convenient to introduce $\sigma_I$ that defines the 
standard deviation for the first element $I$ of $\mathbf{I}$
after $n_{\rm{ph}}$ one-photon experiments:
\begin{equation}
\sigma_I^2=\frac{1}{n_{\rm{ph}}-1}
(
\frac{1}{n_{\rm{ph}}}
\sum_{i=1}^{n_{\rm{ph}}} I_i^2-
(
\frac{1}{n_{\rm{ph}}} 
\sum_{i=1}^{n_{\rm{ph}}} I_i
)^2
),
\label{varianceI_eq}
\end{equation}
and where $I_i$ is a short form for ${I}(u_i,v_i)$. 
In turn, for $F_I$:
\begin{equation}
\sigma_{F_I}=\frac{\pi}{2}(1+\cos{(\alpha)}) \sigma_{I}.
\label{varianceFI_eq}
\end{equation}
Analogous expressions
can be written for the other elements of $\mathbf{F}$: 
$F_Q$, $F_U$ and $F_V$. 
 
\begin{figure*}
\begin{centering}
\noindent\includegraphics{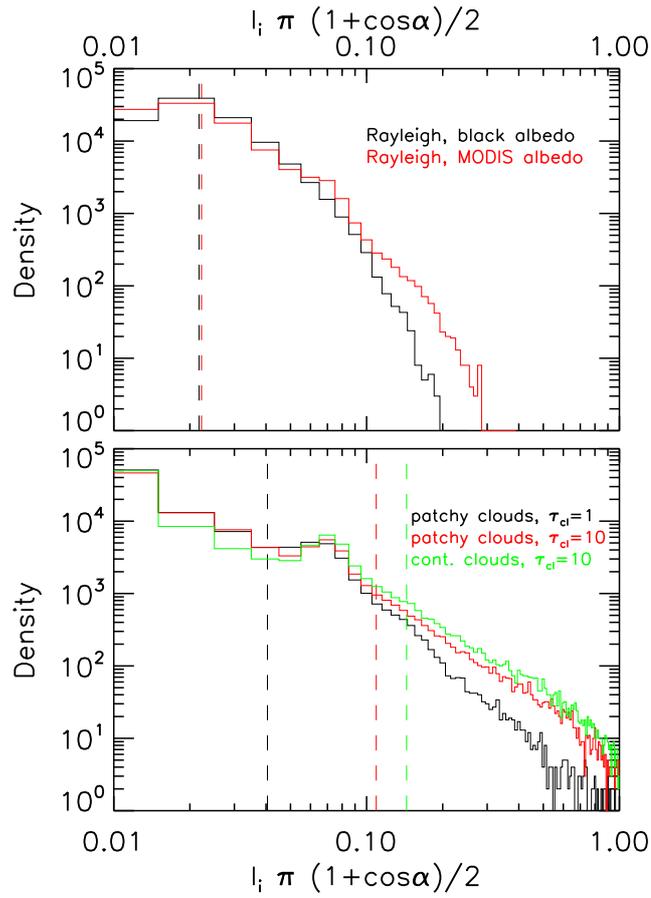}
\caption{\label{histograms_fig} 
Histograms for the partial evaluations of Eq. (\ref{summation_eq})
with the PBMC algorithm and various surface/cloud configurations. 
Each experiment comprises a total of $n_{\rm{ph}}$=10$^5$ one-photon realisations. 
The top and bottom panels are specific to cloud-free and
cloud-covered atmospheres respectively. 
The experiment corresponds to the viewing/illumination geometry of 
Fig. \ref{albedocloudmap_fig}, with $\alpha$=98.7$^{\circ}$. The dashed vertical
lines are mean value estimates for $F_I$, as follows from Eq. (\ref{summation_eq}).
} 
\end{centering}
%/Users/agm121/Desktop/MESSENGERhome/TESTS/MC_EARTHDISK_30SLABS/TOOLS/histoplot.pro
\end{figure*}

Figure (\ref{histograms_fig}) shows histograms of the
$I_i \pi(1+\cos{(\alpha)})/2$ values obtained in the application of
Eq. (\ref{summation_eq}) to one of the atmospheric configurations discussed 
in \S\ref{messenger_section}. 
The top panel corresponds to Rayleigh atmospheres at 0.48 $\mu$m 
above a black surface (black curve) or above the non-uniform MODIS albedo 
map (red curve). 
A non-uniform surface albedo leads to a slightly broader histogram and a correspondingly
larger value of $\sigma_{F_I}$/$F_I$, as shown in Table (\ref{sigma_table}). 
$\sigma_{F_I}$/$F_I$ is indeed a measure of the accuracy associated 
with the estimated $F_I$ after $n_{\rm{ph}}$ one-photon experiments. 
 
The bottom panel of Fig. (\ref{histograms_fig}) shows the histograms obtained 
for configurations that in addition incorporate the MODIS cloud map 
(black and red curves, $\tau_c$=1 and 10 respectively) or a continuous cloud 
cover of optical thickness $\tau_c$=10 (green curve). 
From the comparison of the two panels, it becomes clear that clouds 
broaden the possible outcome from the single-photon experiments, which 
affects the accuracy of the algorithm for a prescribed number of one-photon 
experiments.
From the values of $\sigma_{F_I}$/$F_I$ 
and computational times for $n_{\rm{ph}}$=10$^5$ 
listed in Table (\ref{sigma_table}), it is straightforward to estimate the 
accuracy and computational time for an arbitrary  $n_{\rm{ph}}$. 
In all cases, estimates of $F_I$ accurate to within 3\% (1\%) can be obtained 
in about 1 (10) minutes. 
A clear advantage of the current approach is that there is little or no overhead associated
with the integration over a non-uniform disk, as shown in Table (\ref{sigma_table}).

%\clearpage

\subsection{Earth simulations from 0.47 to 2.13 $\mu$m}

\subsubsection{\label{deglinpol_sec}Brightness and degree of linear polarisation}

The number of possibilities to explore in terms of viewing geometries and
surface/atmosphere configurations is infinite. Important conclusions 
for configurations that include polarisation have 
been presented in the literature
\citep[e.g.][]{bailey2007,karalidietal2011,karalidietal2012,karalidistam2012,stam2008,zuggeretal2010,zuggeretal2011}.
To demonstrate the PBMC algorithm, here I will focus
on the impact of the cloud optical thickness
and adopt $\tau_{\rm{c}}$=0 and 5 for two separate sets of calculations.
It is assumed that 
the planet follows an edge-on circular orbit with the sub-observer point
permanently on the equator of the planet.  
Because with the present definition of $F_Q$ and $F_U$, the relation 
$|$$F_U$$|$$<<$$|$$F_Q$$|$ generally holds (see Section \S\ref{deglinpol_sec}), 
only $F_I$ and $F_Q$/$F_I$ are considered here.

\begin{figure*}
\begin{centering}
\noindent\includegraphics[width=15cm]{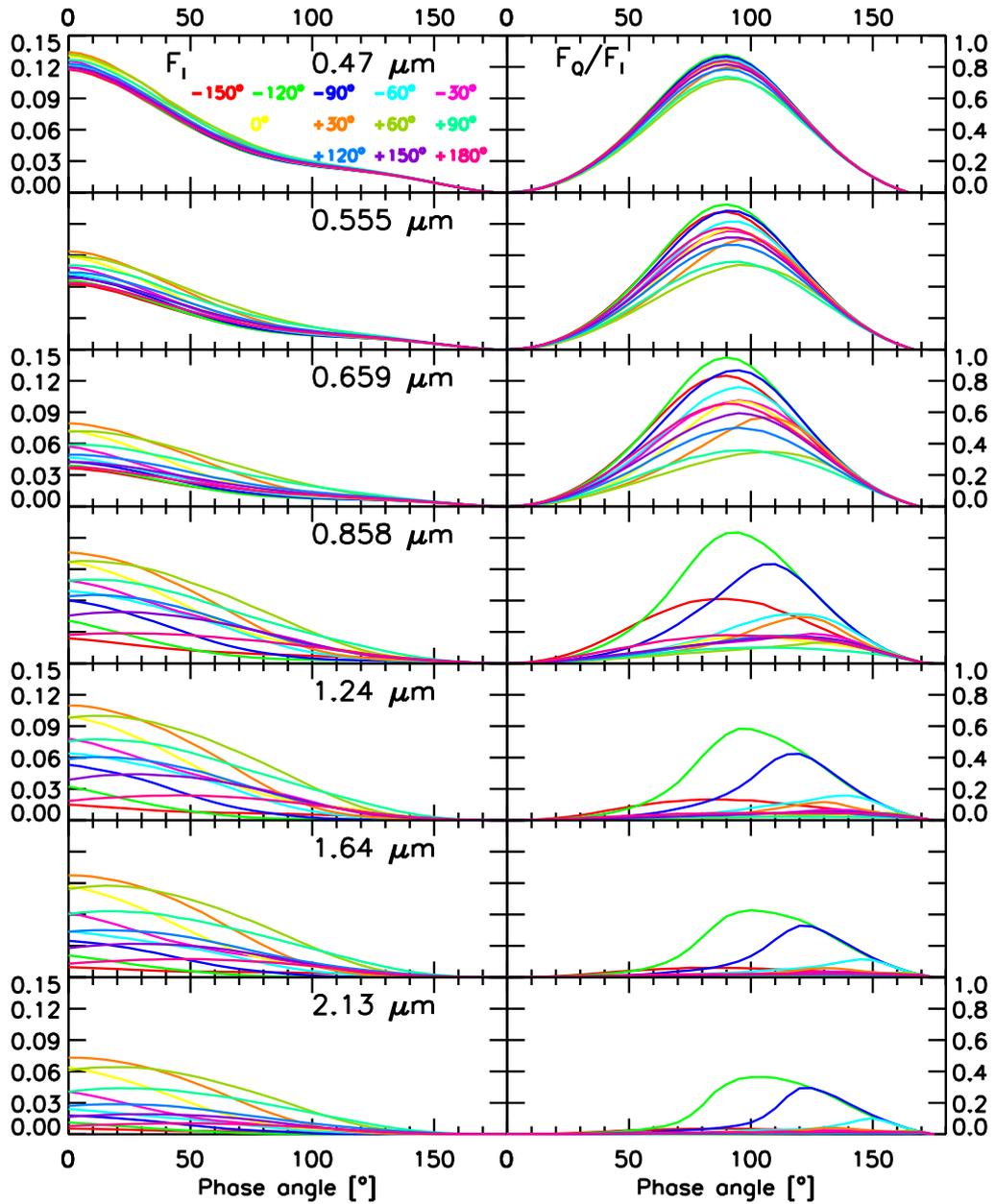} 
\caption{Brightness (left) and polarisation (right) curves for cloud-free
conditions. Only the $\alpha$=0--180$^{\circ}$ range is presented.  
From top to bottom, wavelengths from 0.47 to 2.13 $\mu$m. Within
each panel, the different colour curves refer to different sub-observer
longitudes. }
\label{phasecurveR_fig}
\end{centering}
\end{figure*}

\begin{figure*}
\begin{centering}
\noindent\includegraphics[width=15cm]{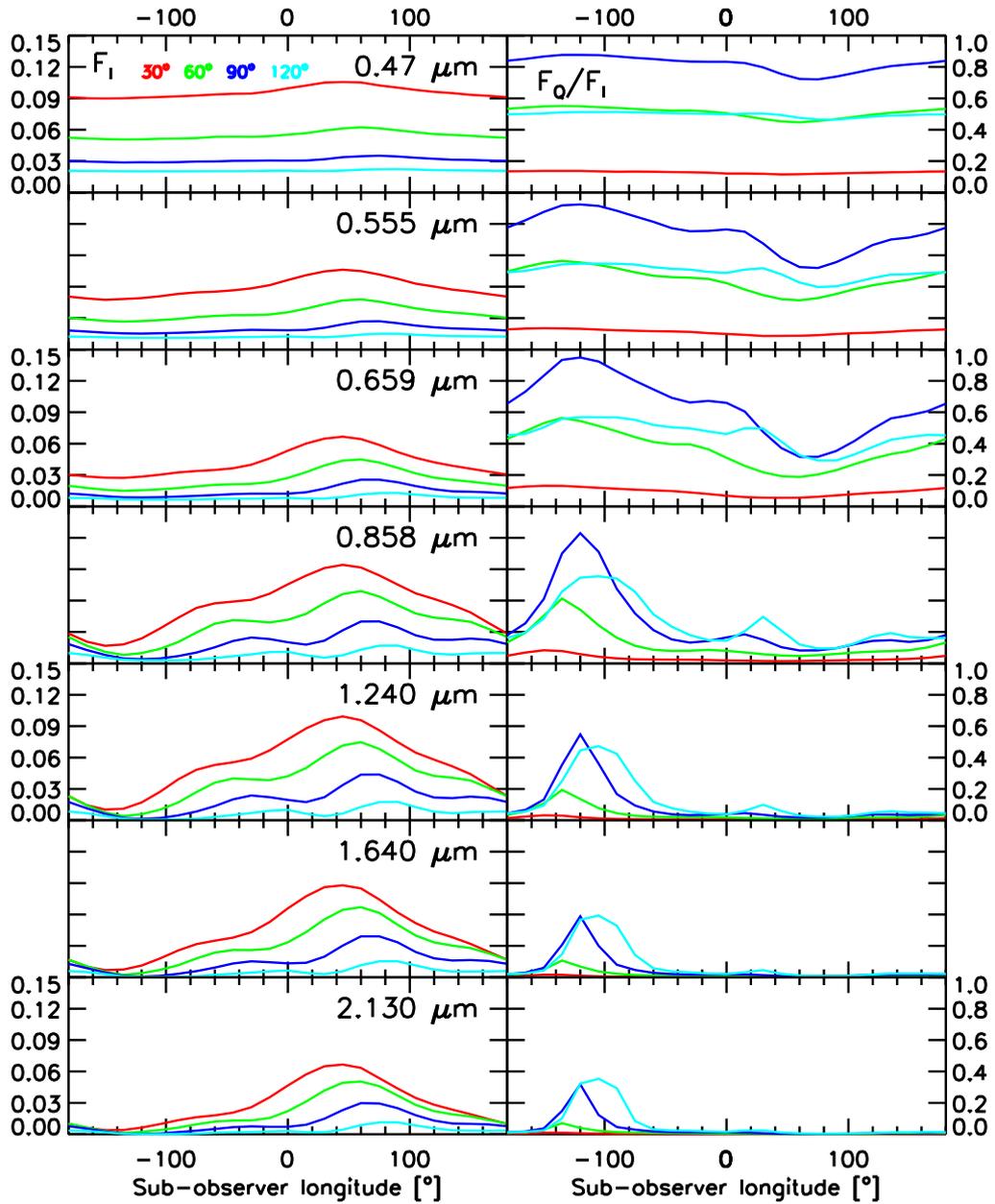} 
\caption{
Brightness (left) and polarisation (right) curves for cloud-free
conditions.
\textbf{Equivalent} to Fig. (\ref{phasecurveC_fig}) but substantiating modulations with
sub-observer longitude or, equivalently, with local time. 
Within each panel, the different colour curves refer to different phase angles. 
}
\label{diurnalcurveR_fig}
\end{centering}
\end{figure*}

Figure (\ref{phasecurveR_fig}) corresponds to $\tau_{\rm{c}}$=0 
(brightness $F_I$ on the left and  polarisation $F_Q$/$F_I$ on the right)
and shows essentially Rayleigh-scattering phase curves \citep{buenzlischmid2009}. 
The cloud-free calculations were carried out with $n_{\rm{ph}}$=10$^5$, which entails
that the results are accurate to better than 1\%.
For each wavelength, the different colour curves correspond to different 
sub-observer longitudes. Local changes in the albedo (and in particular, whether
land/ocean dominate the field of view) modulate both the brightness
and polarisation of the planet. 
The modulation is stronger at the longer
wavelengths because the Rayleigh optical thickness drops rapidly with wavelength. 
Figure (\ref{diurnalcurveR_fig}) substantiates the dependence of 
$F_I$ and $F_Q$/$F_I$ on the sub-observer longitude, or equivalently,  
on local time. 
%provided that the planet orbital period is much longer than its rotation period. 

\begin{figure*}
\begin{centering}
\noindent\includegraphics[width=15cm]{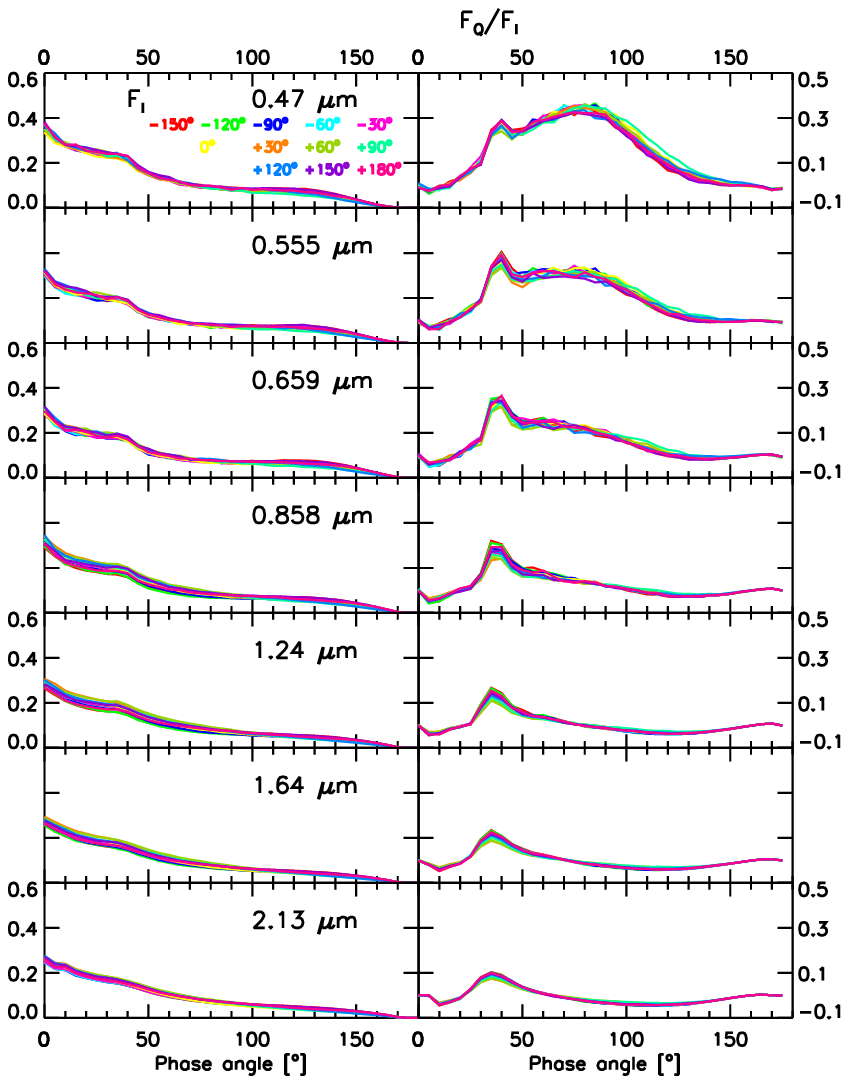} 
\caption{Equivalent to Fig. (\ref{phasecurveR_fig}) but for an atmosphere
partially covered by clouds of optical thickness $\tau_c$=5. 
The estimated relative accuracy is 3\% and 6\% for brightness and
polarisation.
}
\label{phasecurveC_fig}
\end{centering}
\end{figure*}

\begin{figure*}
\begin{centering}
\noindent\includegraphics[width=15cm]{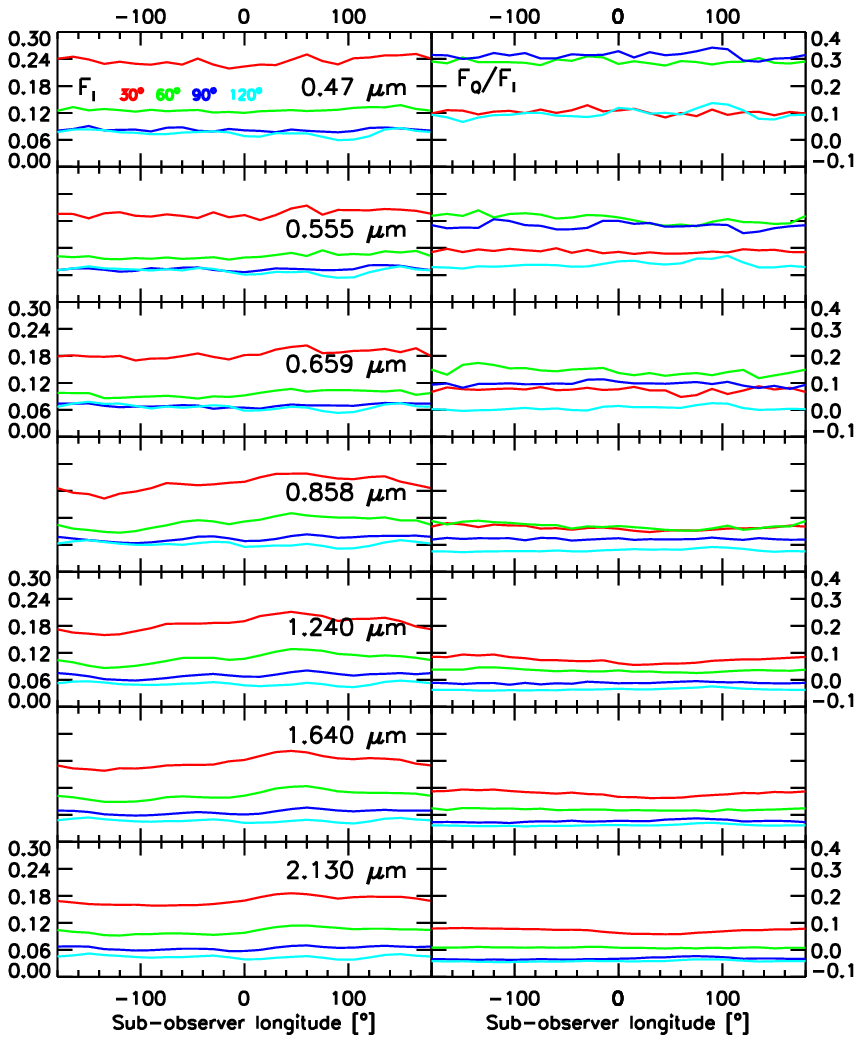} 
\caption{Equivalent to Fig. (\ref{diurnalcurveR_fig}) but for an atmosphere
partially covered by clouds of optical thickness $\tau_c$=5. 
The estimated relative accuracy is 3\% and 6\% for brightness and
polarisation.
}
\label{diurnalcurveC_fig}
\end{centering}
\end{figure*}

Figures (\ref{phasecurveC_fig})--(\ref{diurnalcurveC_fig}) are analogous 
to Figs. (\ref{phasecurveR_fig})--(\ref{diurnalcurveR_fig}) but incorporate 
the MODIS map of cloud fractions for 3rd August 2005, clouds located at 
2--3 km altitude and optical thickness $\tau_c$$\sim$5. 
The implemented cloud droplets' scattering cross sections vary by less
than 10\% from 0.47 to 2.13 $\mu$m, which justifies the adoption of a wavelength-independent
optical thickness for the clouds.
The calculations were carried out
with $n_{\rm{ph}}$=10$^6$ to ensure relative accuracies 
$\sigma_{F_I}$/$F_I$$\sim$3\%.
In the comparison between the cloud-free and cloud-covered configurations, 
two basic conclusions emerge. First, the curves for $\tau_{\rm{c}}$=5 
show a dependence with phase angle that reveals some basic properties of the 
scattering cloud particles. Indeed, the polarisation phase curves show a peak 
at $\alpha$=30--40$^{\circ}$ due to the primary rainbow of water droplets
\citep{bailey2007,karalidietal2011,karalidietal2012,stam2008}. 
Second, clouds attenuate much of the diurnal variability and, 
typically, the diurnal modulation in brightness is easier to pick up than the
modulation in polarisation.

%Thus, brightness measurements appear to be better suited in the investigation of
%diurnal changes in a planet's cover, whereas polarisation yields a better insight 
%if measurements over a broad range of phase angles are available.

\citet{stam2008} and follow-up work \citep{karalidistam2012,karalidietal2011,karalidietal2012}
have explored the appearance in both brightness and polarisation of Earth-like exoplanets. 
\citet{stam2008} and \citet{karalidietal2011} introduce the concept of quasi-horizontally
inhomogeneous planets, which allows them to estimate the outgoing radiation from a
horizontally non-uniform planet as a weighted sum of the signal from uniform-planet solutions. 
Their weighting scheme is based on the fractional coverage of each uniform 
configuration. 
\citet{karalidistam2012} and \citet{karalidietal2012}, however, utilise a fully 
inhomogeneous treatment of the planet.

Comparison of Fig. 13 of \citet{stam2008} at wavelengths of 0.44 and 0.87 $\mu$m
and Fig. (\ref{diurnalcurveR_fig}) here 
at wavelengths of 0.47 and 0.858 $\mu$m is relevant. For cloud-free conditions and a 
phase angle $\alpha$=90$^{\circ}$, $F_I$$\sim$0.03--0.04 and $F_Q$/$F_I$$\sim$0.8
at the shorter wavelength in both works. 
At the longer wavelength, also for $\alpha$=90$^{\circ}$, \citet{stam2008} reports three
distinct peaks at different sub-observer longitudes and highest $F_I$ values of 0.06. 
The three-peak structure is reproduced by the calculations presented here, but the 
highest $F_I$ here is 0.045, somewhat lower than 0.06 in \citet{stam2008}. 
The different treatment of the albedo 
in the two works, 
in particular their magnitudes and the absence of Fresnel reflection in the current one,
 is a likely reason for the discrepancy. 
In polarisation, both works show clear structure that correlates inversely with the brightness. 
The intensity of the $F_Q$/$F_I$ peaks, however, differs between the two works. 
Since high $F_Q$/$F_I$ values typically match dark areas of the planet, again, 
the discrepancies between the two works probably arise from the different 
treatment of the surface albedo.

Similarly, 
comparison of Fig. 14 of \citet{stam2008} and Fig. (\ref{diurnalcurveC_fig}) here reveals
information about the role of clouds. It is difficult to draw definite conclusions from 
the comparison between the two sets of curves because the optical thickness and droplet size of the
clouds implemented in each work are different. The two sets show, however, that the diurnal variability of 
both $F_I$ and $F_Q$/$F_I$ is highly attenuated at the shorter wavelength, but that the 
variability in $F_I$  is still distinct at the longer one.

%I note, however, that in the framework of the PBMC approach, there is no significant 
%computational penalty in treating the planet as horizontally non-uniform, as 
%discussed earlier and summarised in Table (\ref{sigma_table}).

\citet{karalidistam2012} and \citet{karalidietal2012} investigate further the
simulated signal from Earth-like exoplanets and take into account non-uniform cloud and albedo
properties. \citet{karalidietal2012} consider a realistic cloud map based on
MODIS data. Their work emphasises the detectability of features like the primary rainbow, 
which originates from Mie scattering in spherical droplets but may be masked by scattering
in ice particles; and, the impact of non-uniformities on the disk-integrated 
signal. A case-by-case comparison between those works and the results presented in Figs.
(\ref{phasecurveR_fig})--(\ref{diurnalcurveC_fig}) here is not feasible. 
It is worth noting, nevertheless, that
the main features that appear in the curves of Figs. 
(\ref{phasecurveC_fig})--(\ref{diurnalcurveC_fig}) here are also found in Figs. 6--8 of
\citet{karalidietal2012}. This includes the occurrence of the primary rainbow in 
both $F_I$ and $F_Q$/$F_I$. 
The significant variety in the curves reported in those works and here illustrates 
the complexity of the problem.  
It also cautions against quick conclusions when the time comes that either brightness
or polarisation curves of Earth-like exoplanets become available.

The near infrared 
(e.g. 1.55--1.75; 2.1--2.3 $\mu$m) 
has been proposed as a better option than visible wavelengths
in the characterisation of surface features that might occur at specific local times 
(high/low reflecting surfaces such as desert/ocean, respectively) 
and phase angles (specular reflection from an ocean) \citep{zuggeretal2011}. 
The simulations here (see curves for $\lambda$$>$1 $\mu$m in Figs. \ref{phasecurveR_fig} and \ref{phasecurveC_fig})
confirm (compare to Fig. 3 in \citet{zuggeretal2011})
that contrasts in the near infrared can be
high for cloud-free conditions, but become less distinct at moderate cloud opacities. 
Since the current implementation of the PBMC 
model does only account for Lambert reflection at the surface, a more direct comparison 
with \citet{zuggeretal2011} is not immediately possible.

%\clearpage
\subsubsection{\label{anglepolarisation_sec}Angle of linear polarisation}

Non-uniform surface albedos  and patchy clouds
introduce an asymmetry between the northern and southern hemispheres that may lead
to non-zero values of $F_U$. The angle of polarisation $\chi$, 
defined through $\tan{2\chi}$=$F_U/F_Q$, 
expresses the orientation of the outgoing 
polarisation vector with respect to the reference plane.
The reference plane is normal to the planet scattering plane; 
the latter is formed by the viewing and illumination directions. 
Because each photon is tracked with respect to the same three-dimensional absolute 
rest frame and the same reference plane is utilised to express the Stokes vector, 
all the photons' Stokes vectors can be added without further manipulation.
The model calculations presented above show, however, that the hemispherically-averaged $U$ components take 
different signs at the northern and southern hemispheres and that
summation over the entire disk leads to the effective cancellation of $F_U$, i.e.
$|F_U$/$F_Q|$$<<$1 away from the neutral points where $F_Q$$\approx$0.

\begin{figure*}
\begin{centering}
\noindent\includegraphics{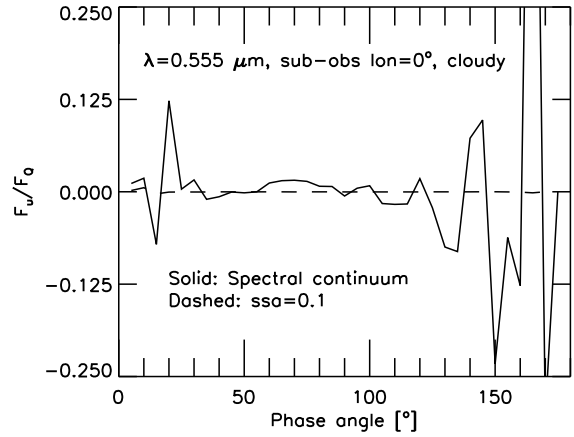}
\caption{\label{rotation_fig} 
Solid curve: Ratio $F_U$/$F_Q$ versus phase angle for the cloudy configuration described in
Fig. (\ref{phasecurveC_fig}) at a wavelength of 0.555 $\mu$m and a sub-observer longitude of 0$^{\circ}$.
The ratio remains close to zero for most phase angles; deviations to that trend 
are largely due to $F_Q$ values nearing zero. 
Dashed curve: Ratio $F_U$/$F_Q$ in the same configuration 
but with a prescribed atmospheric single scattering albedo (ssa$\equiv$$\varpi$) of 0.1
at all altitudes. By minimising the significance of multiple scattering,
the dashed curve provides a better insight into conditions within a strong molecular 
absorption band. 
} 
\end{centering}
%/Users/agm121/Desktop/MESSENGERhome/MC_EARTHDISK_MESS/TOOLS/alphatime.pro
\end{figure*}

For reference, Fig. (\ref{rotation_fig}; solid curve) shows the change in the ratio 
$F_U$/$F_Q$ with phase angle at a wavelength 
of 0.555 $\mu$m and sub-observer longitude of 0$^{\circ}$ in one of the cloud-covered configurations
investigated in Fig. (\ref{phasecurveC_fig}).
The ratio $F_U$/$F_Q$ remains small at all phase angles, with deviations to that trend
arising where $F_Q$ becomes small and the statistical errors large.
Thus, the planet's polarisation vector is 
preferentially aligned with the planet scattering plane
(if $F_Q$$<$0) or with the reference plane (if $F_Q$$>$0), 
also in the presence of non-uniformities at both the surface and cloud levels.

It is possible to assess what occurs within a molecular absorption band by 
setting the single scattering albedo in the atmospheric model $\varpi$$<$1. 
The dashed curve in Fig. (\ref{rotation_fig}) presents the ratio $F_U$/$F_Q$ for an
adopted $\varpi$=0.1 at all altitudes. In those conditions, single scattering 
dominates and the ratio $F_U$/$F_Q$ becomes smaller.
According to the calculations shown in Fig. (\ref{rotation_fig}), 
the angle of linear polarisation 
$\chi$ is either close to zero or to $\pi$/2 
both within and outside molecular absorption bands.

%\clearpage
\subsubsection{Circular polarisation}

Circular polarisation in the light scattered from planetary atmospheres typically 
results from two
or more collisions of the photons within the medium.
In the cases that measurements have been attempted \citep[e.g.][]{kempetal1971,sparksetal2005,swedlundetal1972}, 
the degree of circular polarisation is orders of magnitude
smaller than for linear polarisation, which entails that 
observations of circular polarisation are always challenging. 
%In addition, 
%the theoretical understanding provided by the 
%measurements is not always immediately clear \citep{}. 
%These issues may help understand the relative paucity of observational and theoretical work on the topic. 

Unlike abiotic material, living organisms may lead to distinct circular 
polarisation signatures \citep[e.g.][]{sparksetal2009}
potentially amenable to remote sensing. 
Indeed, 
the prospects of using circular polarisation as a biomarker in the investigation of
habitable exoplanets has prompted new research on this front
\citep[e.g.][]{nagdimunovetal2014,sparksetal2009}.

\begin{figure*}
\begin{centering}
\noindent\includegraphics[width=9cm]{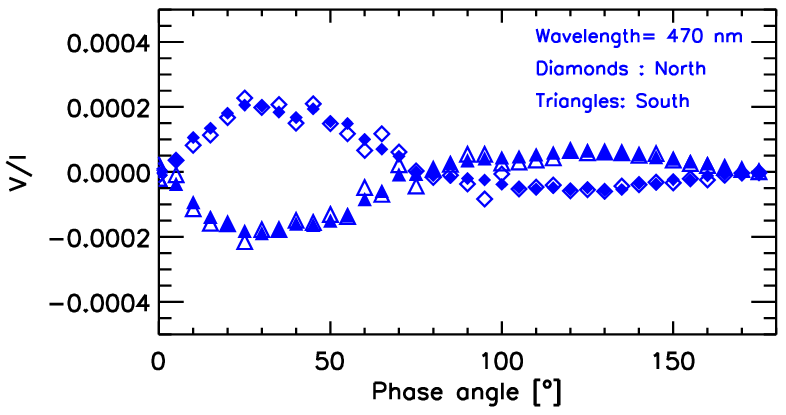}
\\
\vspace{-0.60cm} 
\noindent\includegraphics[width=9cm]{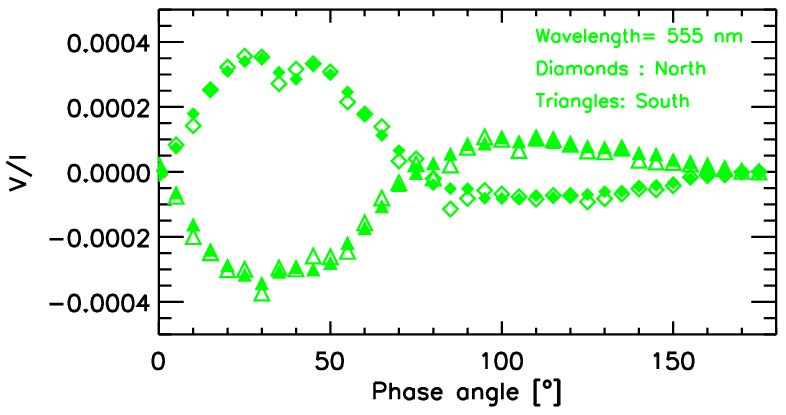}
\\
\vspace{-0.60cm} 
\noindent\includegraphics[width=9cm]{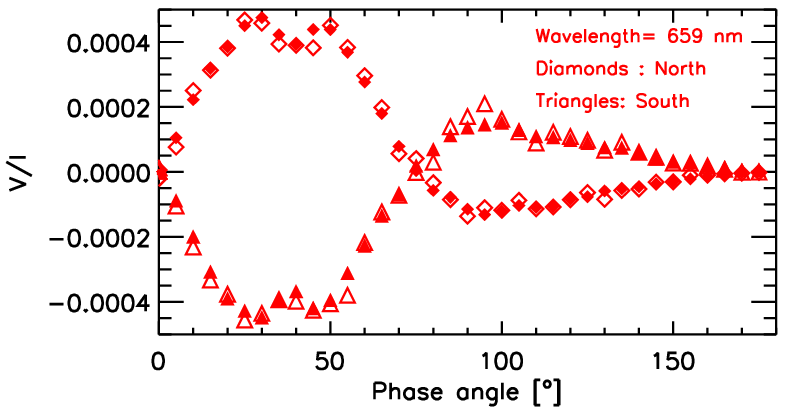}
\caption{\label{LC_VI_fig} Degree of circular polarisation V/I for the cloudy configuration
described in Fig. (\ref{phasecurveC_fig}) at the three 
specified wavelengths and a sub-observer longitude of
0$^{\circ}$. Separate curves are shown for calculations over the northern and southern 
hemispheres. Open and filled symbols correspond to calculations with
$n_{\rm{ph}}$ of about 10$^9$ and 10$^{10}$, respectively. 
} 
\end{centering}
\end{figure*}

The Earth model considered here does not include circular polarisation 
effects associated with living organisms. It is relevant, nevertheless, 
to assess the suitability of the PBMC algorithm for a prospective investigation 
of biogenic circular polarisation. 
On the basis of the configurations investigated in Fig. 
(\ref{phasecurveC_fig}) for cloudy atmospheres, 
I have calculated $F_V$/$F_I$ vs. phase angle at the wavelengths of 0.470, 0.555 and 0.659 $\mu$m and a
sub-observer longitude of 0$^{\circ}$. 
The calculations are presented in Fig. (\ref{LC_VI_fig}) separately for the northern and 
southern hemispheres, and with $n_{\rm{ph}}$=10$^9$ and 10$^{10}$ one-photon 
experiments per phase angle.  
As expected \citep{kawata1978,kempetal1971}, 
the two hemispheric components have comparable magnitudes but opposite signs, 
and tend to cancel out when the integration is extended over the entire planet disk. 
The numerical experiments suggest that a number 
$n_{\rm{ph}}$=10$^9$ of one-photon numerical experiments are needed
to reach acceptable accuracies. This is about three orders of magnitude more than 
the needs for linear polarisation, which 
conveys one of the difficulties in predicting and interpreting
circular polarisation features.

\clearpage
\section{\label{messenger_section}Messenger diurnal light curves}

%NASA's Messenger spacecraft was launched in August 2004. 
On cruise towards Mercury, the NASA Messenger spacecraft performed an Earth flyby
in August 2005 \citep{mcnuttetal2008}. 
Observations of Earth during the gravity assist include imagery with the
Mercury Dual Imaging System (MDIS) cameras.  
Of particular interest here are the colour sequences of full-disk images 
captured with the Wide Angle Camera (WAC) on the departure leg.
Each sequence comprises nearly-simultaneous observations over a narrow-band
filter of band center/width ($\mu$m), $\lambda$/$\Delta$$\lambda$= 0.480/0.01, 
0.560/0.006 and 0.63/0.0055. 
With a cadence of 3 images every 4 minutes, the total number of
images amounts to 1080. 

\begin{figure*}
\begin{centering}
\noindent\includegraphics{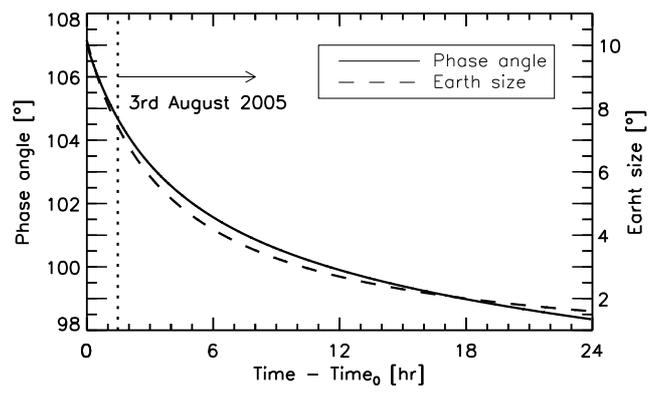}
%\\
%\vspace{-1.00cm} 
%\noindent\includegraphics{./FIGURES/LCMESS1b.eps}
\caption{\label{alphatime_fig} Sun-Earth centre-spacecraft  
phase angle and apparent angular size of Earth during the Messenger flyby.
} 
\end{centering}
%/Users/agm121/Desktop/MESSENGERhome/MC_EARTHDISK_MESS/TOOLS/alphatime.pro
\end{figure*}

As the spacecraft departs, 
the Earth angular diameter evolves from 10.2$^{\circ}$ 
(nearly filling the WAC field of view of 10.5$^{\circ}$) 
at the start of the sequence to 1.6$^{\circ}$ at the end of the sequence 24
hours later. At its farthest during the sequence (about 457,000 km from Earth), 
the spacecraft is beyond the Moon orbit. 
Correspondingly, the Sun-planet centre-spacecraft
phase angle $\alpha$ varies from 107$^{\circ}$ to 98$^{\circ}$. 
Figure (\ref{alphatime_fig}) presents the two geometric parameters against time
since the start of the sequence on 2nd August at about 22:31UT. 
%Figure (\ref{}) shows a snapshot of Earth as viewed from Messenger at the end of the
%sequence, together with the albedo map at 0.47 $\mu$m implemented in the
%calculations and the corresponding
%distribution of clouds on 3rd August.

The WAC images were downloaded from the Planetary Data System and read 
with the software made available at the Small Bodies Node.\footnote{ 
http:$//$pdssbn.astro.umd.edu$/$tools$/$tools$\_$readPDS.shtml}
The basic treatment of the images is similar to that
outlined by \citet{domingueetal2010} in their analysis of the 
whole-disk optical properties of Mercury. 
Integration of the planet brightness was conducted over a rectangular box
that contains the Earth visible disk. 
Residual background values were estimated from two stripes running
north-south on each side of the box. When possible, the box was designed
to extend beyond the visible edge by 0.2--0.3 
Earth radii, and the adjacent stripes were 10 pixel wide. 
Other combinations of box size and stripe width resulted in very similar
conclusions. 
The corrected brightness was then summed over the 
integration box and multiplied by the solid angle per pixel to yield the planet's
irradiance as measured from the spacecraft vantage point, $F_p$. 
To derive the planet phase function $A_g$$\Phi$, I applied the equation:
\begin{equation}
F_p=  \left( \frac{R_p}{\Delta} \right)^2  \left( \frac{1}{a_{\star} [\mbox{AU}]} \right)^2
A_g \Phi
F_{\star},
\end{equation}
where $R_p$ is the Earth's mean radius (=6371 km), 
$\Delta$ is the spacecraft-to-Earth centre distance, 
and $a_{\star}$ is the Earth's orbital distance. 
$F_{\star}$ is the solar irradiance at 1 Astronomical Unit specific to the camera filter
\citep{micketal2012}. 
A related magnitude, the apparent albedo, can be obtained from: 
\begin{equation}
A_{\rm{app}}=\frac{A_g \Phi(\alpha)}{(A_g \Phi(\alpha))_{\rm{Lambert}}}=
\frac{A_g \Phi(\alpha)}{\frac{2}{3} \frac{1}{\pi} [\sin{\alpha}+(\pi-\alpha)\cos{\alpha}]}, 
\end{equation}
that refers $A_g \Phi(\alpha)$ to that for a Lambert sphere. 

\begin{figure*}
\begin{centering}
\noindent\includegraphics[width=9cm]{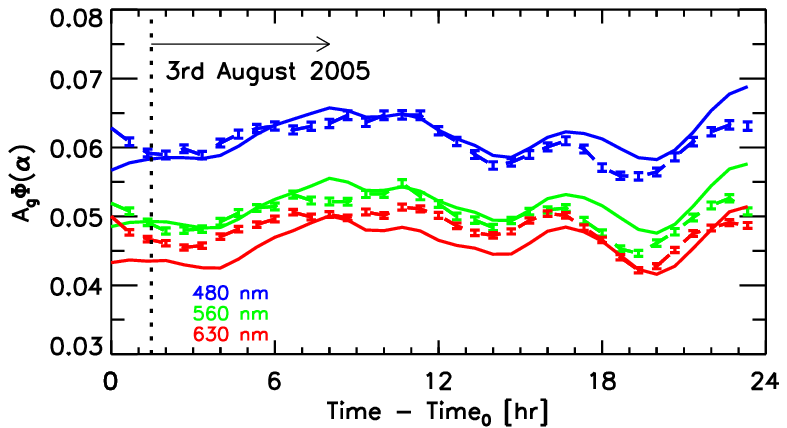} 
%\\
%\vspace{-0.50cm} 
%\noindent\includegraphics[width=9cm]{./FIGURES/LCMESS1b.eps}
\\
\vspace{-0.60cm} 
\noindent\includegraphics[width=9cm]{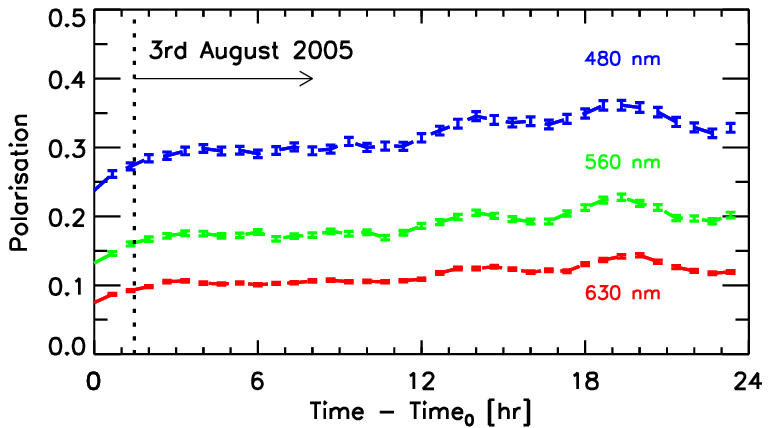}
\\
\vspace{-1.30cm} 
\noindent\includegraphics[width=9cm]{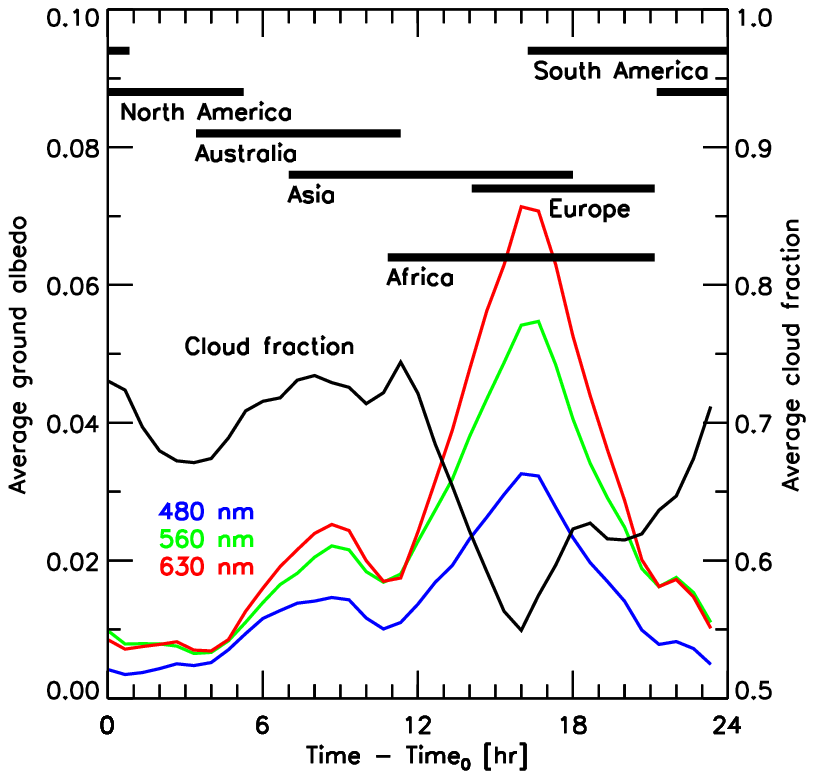}
\caption{\label{LCMESS1_fig} 
Top panel: Messenger diurnal light curves. 
Solid curves are Messenger data; dashed curves are model
predictions with clouds at 2--3 km altitude and 
cloud optical thickness $\tau_c$=3. For the model predictions, 
the error bars are standard deviations    
$\sigma_{F_I}$/$F_I${$\sim$}1\%.
Middle panel: Model-predicted polarisation light curves for the 
configuration of the top panel.
Bottom panel: Average cloud fraction and surface albedos over the Earth visible disk
at each instant during the observing sequence.
} 
\end{centering}
%/Users/agarcia/MESSENGER/MC_EARTHDISK_MESS/TOOLS/lcmess1.pro
\end{figure*}

Figure (\ref{LCMESS1_fig}) shows the Messenger $A_g$$\Phi$ 
light curves  (top panel; solid)  against time for each colour filter. 
For comparison, \citet{mallama2009} reports $A_g$$\Phi$(90$^{\circ}$)=0.06 
and $A_g$$\Phi$(120$^{\circ}$)=0.05 from earthshine observations in broadband
visible light \citep{goodeetal2001}. 
The Messenger curves exhibit variations of peak-to-peak amplitude up to 
$\sim$20\% over times of hours, which is consistent with the findings by EPOXI 
\citep{livengoodetal2011}.

The evolving phase angle precludes an interpretation 
of the light curves based solely on diurnal changes in the Earth cover. 
A detailed observation-model analysis is deferred to future work. 
It is worth noting, however, 
that the model captures the major features in the Messenger light curves. 
Using the atmosphere/surface prescriptions of \S\ref{atmosphere_section} 
(and in particular, the albedo maps at 0.47, 0.555 and 0.658 $\mu$m, 
the latter being somewhat off-set from the MDIS/WAC band centre), 
the Messenger $A_g$$\Phi$ points are reasonably 
well reproduced by the model calculations (Fig. \ref{LCMESS1_fig}; top panel; dashed)
that incorporate the MODIS cloud map specific to 3rd August 2005, clouds located at 
2--3 km altitude and optical thickness $\tau_c$$\sim$3, 
and the relevant viewing/illumination geometry. 
The space of model parameters is vast and 
other atmospheric configurations can plausibly lead to similar or better fits. 
At the relevant phase angles, the model predicts strong polarisation at all
three wavelengths, as shown in Fig. (\ref{LCMESS1_fig}; middle panel). 
The model calculations were made with $n_{\rm{ph}}$=10$^7$, 
which entails standard deviations 
$\sigma_{F_I}$/$F_I$$\sim$$\sigma_{F_Q}$/$F_Q$$\sim$1\%.  
The bottom panel of Fig. (\ref{LCMESS1_fig}) shows the corresponding cloud fraction and
surface albedos averaged over the Earth visible disk at each instant during the 
observing sequence. 
The drop in cloudiness and increase in surface albedo
that occurs half way through the observing sequence, especially at the longer wavelengths,
reveals the presence of Africa's vast desert expanses.

\begin{figure*}
\begin{centering}
\noindent\includegraphics[width=9cm]{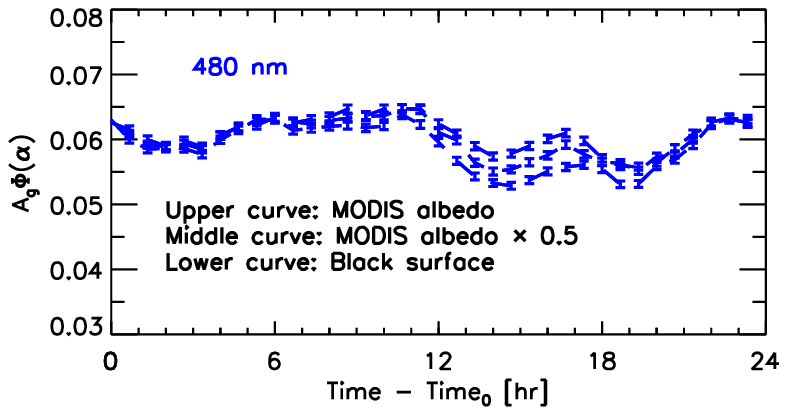} 
\\
\vspace{-0.80cm} 
\noindent\includegraphics[width=9cm]{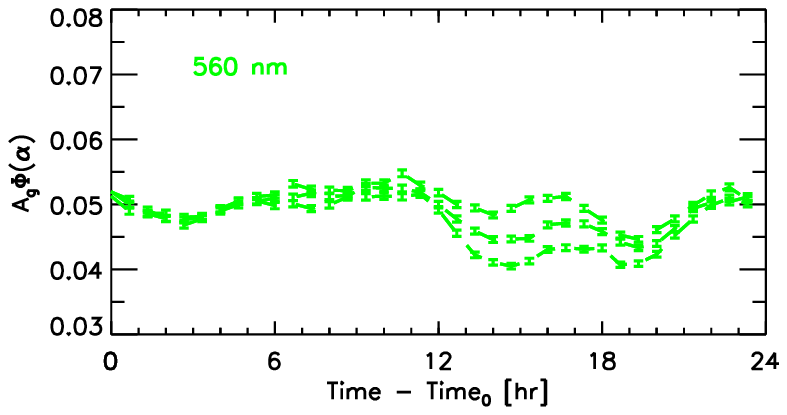} 
\\
\vspace{-0.80cm} 
\noindent\includegraphics[width=9cm]{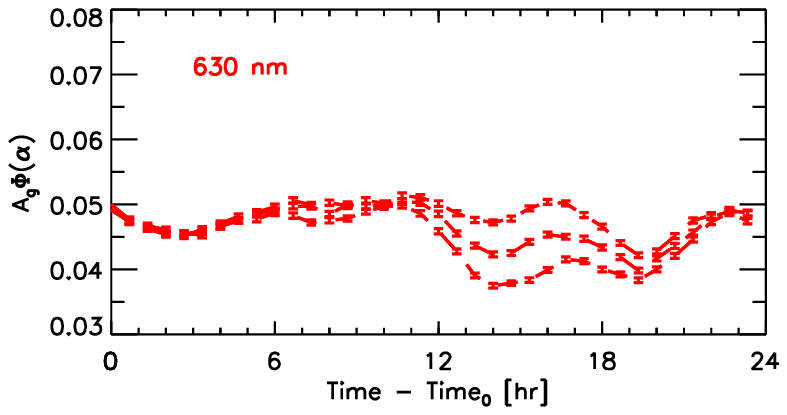} 
\caption{\label{LCMESSpert_fig} 
Model-predicted brightness light curves for conditions specific to the 
Messenger observing sequence. 
The three curves in each panel represent three differently scaled versions of the surface albedo.
The surface becomes more apparent between 12 and 20 hours, with Africa well 
into the field of view, because the cloud fraction is 
lower and the surface albedo is higher. The contrast is higher at 630 nm than at either
480 or 560 nm because Rayleigh scattering becomes less efficient at longer wavelengths 
and more photons penetrate deeper into the atmosphere. 
} 
\end{centering}
\end{figure*}

In preparation of a more thorough exploration, Fig. (\ref{LCMESSpert_fig}) investigates
the sensitivity of the model-predicted Messenger light curves to the surface albedo. 
Postulated changes in the reflective properties of the surface are particularly clear from 
12 to 20 hours in the observing sequence. This is related to two
different factors, namely the low cloud fraction and high surface albedos at the time. 
Decreased Rayleigh scattering allows photons to penetrate deeper, which leads to 
increased relative contrasts at the longer wavelength. 
The simulation above a black surface clearly demonstrates that most, but not all, 
of the variability in the
light curve at the three wavelengths originates from cloud patchiness.

\clearpage
\section{\label{outlook_section}Summary and outlook}

Models will play key roles in the interpretation of future observations of 
exoplanetary atmospheres. 
Here, I presented a method to predict the disk-integrated Stokes vector 
of stellar radiation reflected from a planet.
The method is flexible and handles variations in the planet 
optical properties in both the vertical and horizontal directions.
A major advantage of this approach is that its computational cost is not significantly
affected by non-uniformities at the atmospheric and cloud levels.
Being based on Monte Carlo integration, the accuracy (and in
turn, the computational time) of a calculation can be established a priori. 
Typically, solutions 
for a specific viewing/illumination geometry
accurate to within 3\% 
are produced in one minute or less for planet configurations as elaborate
as those described here for an Earth replica. 
In the future, it will be interesting to 
address a detailed analysis of the Messenger light curves and
investigate the complementarity of brightness and polarisation measurements
in the characterisation of Earth-like exoplanets over long time baselines. 
\\

\underline{\textbf{Acknowledgments}}\\

I gratefully acknowledge the assistance of Jake Ritchie 
(University of Maryland, MD) 
with the software for reading the Messenger/MDIS images, and
access to the GeoViz software 
(Henry Throop, http://soc.boulder.swri.edu/nhgv/) 
for the visualization of Earth from the Messenger trajectory.
The Messenger/MDIS images were downloaded from the Planetary Data System. 
I gratefully acknowledge conversations with Tom Enstone (ESA/ESTEC, Netherlands) 
about the Earth simulations.  
Finally, I thank Frank P. Mills (Australian National University, Australia)
for comments on an early version of the manuscript.

%\clearpage
\bibliographystyle{elsarticle-harv}
%\bibliography{<your-bib-database>}

%% Authors are advised to submit their bibtex database files. They are
%% requested to list a bibtex style file in the manuscript if they do
%% not want to use elsarticle-harv.bst.

%% References without bibTeX database:

% \begin{thebibliography}{99}

\newpage

\begin{table}
\caption{ \label{sigma_table} 
Summary of performances of the algorithm for disk integration at 0.48 $\mu$m 
and various Earth configurations discussed in \S\ref{convergence_section}. 
Computational times are based on a 2.8 GHz workstation and $n_{\rm{ph}}$=10$^5$. 
}
\begin{flushleft}
\begin{tabular}{cccc}
\hline
\hline
Surf. albedo & Cloud & $\sigma_{F_I}$/$F_I$$\times$100 & Comp. time [s]\\
\hline
0 & No & 0.24 & 1.5 \\
MODIS & No & 0.33 & 1.4 \\
MODIS & Patchy, $\tau_c$=1 & 7.8 & 2.2 \\
MODIS & Patchy, $\tau_c$=10 & 9.1 & 7.7 \\
MODIS & Continuous, $\tau_c$=10 & 7.7 & 10.4 \\
\hline 
\hline 
\end{tabular} 
\end{flushleft} 
\end{table}

%\begin{figure*}
%\begin{centering}
%\noindent\includegraphics[width=9cm]{./FIGURES/LCMESS2.eps}
%\caption{\label{LCMESS2_fig} 
%Model polarisation curves for the Messenger flyby. 
%Because $\sigma_{F_I}$/$F_I${$\sim$}$\sigma_{F_Q}$/$F_Q$, 
%error bars are estimated as $\sigma_{F_I}$/{$F_I$}{$\times$}2$\sim$2\%.
%} 
%\end{centering}
%%/Users/agm121/Desktop/MESSENGERhome/TESTS/MC_EARTHDISK_30SLABS/TOOLS/histoplot.pro
%\end{figure*} 

\end{document}